\documentclass{PoS}

\usepackage{amsmath}
\usepackage{multirow}
\usepackage{caption}

\usepackage[%
  citestyle=numeric-comp,
  sorting=none,
  backend=bibtex,
  mcite=true,
  eprint=true,
  doi=false,
]{biblatex}
\AtEveryBibitem{\clearfield{title}}
\renewbibmacro{in:}{}
 
\newcommand{\castorc}{Computation-based Science and Technology
  Research Center, The Cyprus Institute, 20 Kavafi Str., Nicosia 2121,
  Cyprus}

\newcommand{\ucy}{Department of Physics, University of Cyprus,
  P.O. Box 20537, 1678 Nicosia, Cyprus}

\newcommand{\desy}{NIC, DESY, Platanenallee 6, D-15738 Zeuthen,
  Germany}

\title{Nucleon electromagnetic form factors from twisted mass lattice QCD}

\ShortTitle{Nucleon EM form factors}

\author{Abdou Abdel-Rehim\\ \castorc \\ E-mail:
  \email{a.abdel-rehim@cyi.ac.cy}}

\author{Constantia Alexandrou\\ \ucy\, and \castorc \\ E-mail:
  \email{alexand@ucy.ac.cy}}

\author{Martha Constantinou\\ \ucy \\ E-mail: \email{marthac@ucy.ac.cy}}

\author{Kyriakos Hadjiyiannakou\\ \ucy \\ E-mail:
  \email{hadjigiannakou.kyriakos@ucy.ac.cy}}

\author{Karl Jansen\\ \desy \\ E-mail:
  \email{karl.jansen@desy.de}}

\author{\speaker{Giannis Koutsou}\\ \castorc
  \\ E-mail:\email{g.koutsou@cyi.ac.cy}}

\abstract{Results on the electromagnetic form factors of the nucleon
  using twisted mass fermion configurations are presented. These
  include a gauge field ensemble simulated with two degenerate light
  quarks yielding a pion mass of around 130~MeV, as well as two
  ensembles that include strange and charm quarks in the sea yielding
  pion masses of 210~MeV and 373~MeV. Details of the methods used and
  systematic errors are discussed, such as noise reduction techniques
  and the effect of excited states contamination.}

\FullConference{The 32nd International Symposium on Lattice Field Theory,\\
		23-28 June, 2014\\
		Columbia University New York, NY}

\begin{document}

\section{Introduction}
The electromagnetic form factors of the proton and neutron are
quantities of continuous interest both experimentally and
theoretically. Electromagnetic scattering of nucleons reveals
properties such as their charge and magnetism, with insight on the
distribution of charge within the nucleon. More recently, precision
experiments have revealed a surprising discrepancy in the charge
radius of the proton. Namely, the proton radius when measured recently
via the Lamb shift of muonic hydrogen~\cite{Pohl:2010zza} has a value
that is smaller by five standard deviations as compared to experiments
using electron scattering and hydrogen Lamb
shift~\cite{Mohr:2012tt}. Lattice QCD can provide insight on the
origin of this discrepancy by evaluating from first principles the QCD
contribution to the proton charge. So far, although consistent across
discretisation schemes, lattice simulations at heavier than physical
pion masses have underestimated the proton
charge~\mcite{Alexandrou:2011db,Alexandrou:2006ru,Syritsyn:2009mx}. With
simulations directly at the physical pion mass now becoming available,
chiral extrapolations are no longer required, thus eliminating one
source of systematic uncertainty and allowing direct contact with
experiment. In this proceedings contribution we present preliminary
results of the nucleon electromagnetic form factors using three
ensembles of twisted mass fermion (TMF) configurations; two $N_{\rm
  f}=2+1+1$ ensembles at pion mass 373~MeV and 210~MeV and one $N_{\rm
  f}=2$ ensemble at the near physical pion mass value of 130~MeV.

\section{Lattice Setup and Methods}
\label{sec:method}
The electromagnetic matrix element of the nucleon can be decomposed
into the Dirac ($F_1$) and Pauli ($F_2$) form factors:
\begin{equation}
\langle N(p',s') | j^\mu | N(p,s) \rangle =
\Big(\frac{m_N^2}{E_N(\mathbf{p}')E_N(\mathbf{p})}\Big)^{\frac{1}{2}}
\bar{u}(p',s')\mathcal{O}^\mu u(p,s),\,\,\,\,\,\,\mathcal{O}^\mu =
\gamma_\mu F_1(q^2) +\frac{i\sigma_{\mu\nu}q^\nu}{2m_N}F_2(q^2),
\end{equation}
with $m_N$ the nucleon mass, $p$ ($s$) and $p'$ ($s'$) the initial and
final momentum (spin) of the nucleon, $u$ are fermion spinors and
$q=p'-p$ is the momentum transfer. Alternatively one can define the
so-called Sachs electric and magnetic form factors:
$G_E(q^2)=F_1(q^2)+ F_2(q^2)$ and $G_M(q^2)=F_1(q^2)+\tau F_2(q^2)$
with $\tau = \frac{q^2}{(2m_N)^2}$. The slope of the form factors at zero
momentum defines the relevant radii, namely the Dirac and Pauli radii
via: $\langle r_i^2\rangle=-\frac{6}{F_i}\frac{dF_i}{dq^2}|_{q^2=0}$
and similarly for the electric and magnetic radii $\langle
r^2_E\rangle$ and $\langle r^2_M\rangle$.

On the lattice, one calculates an appropriate three-point correlation function:
\begin{equation}
G^\mu(\Gamma;\mathbf{q};t_s,t_i) = \sum_{\mathbf{x}_s\mathbf{x}_i}e^{-i\mathbf{p}'\mathbf{x}_s}e^{-i(\mathbf{p}'-\mathbf{p})\mathbf{x}_i}\Gamma^{\alpha\beta}\langle \bar{\chi}^\beta_N(\mathbf{x}_s;t_s) | j^\mu(\mathbf{x}_i;t_i) | \chi^\alpha_N(\mathbf{x}_0;t_0)\rangle
\end{equation}
with $\chi_N$ the nucleon interpolating operators, $x_s$, $x_i$ and
$x_0$ the final (sink), insertion and initial (source) coordinates and
the local electromagnetic current: $j^\mu(x) =
\frac{2}{3}\bar{u}(x)\gamma^\mu u(x)-\frac{1}{3}\bar{d}(x)\gamma^\mu
d(x)$. In this work we use the lattice conserved electromagnetic
current~\cite{Alexandrou:2011db}, and thus lattice results need no
renormalisation. In our setup we fix $\mathbf{p'}=0$ and carry out
the sum over $\mathbf{x}_s$ using a sequential inversion through the
sink. For this we require $t_s-t_0$ and the projection matrix $\Gamma$
to be set before the sequential inversion. The choices
$\Gamma_4=\frac{1}{4}(1+\gamma_4)$ and $\Gamma_k=\sum_j
i\Gamma_4\gamma_5\gamma_j$ isolate $G_E$ and $G_M$ respectively.

\begin{figure}[t]
  \includegraphics[width=\linewidth]{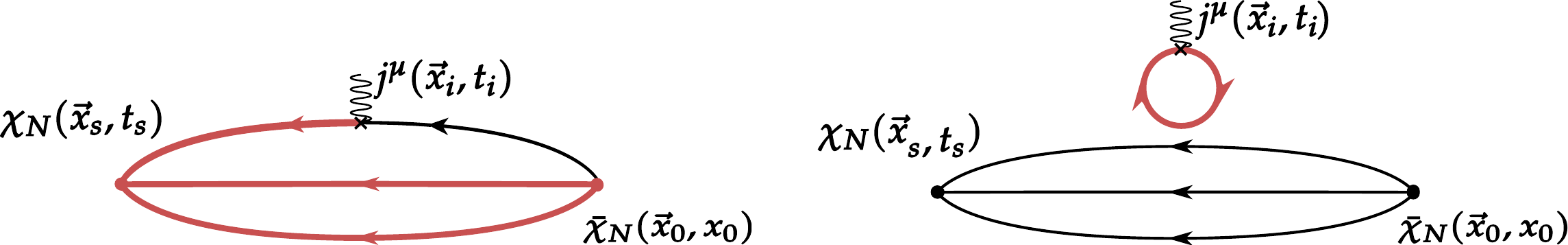}
  \caption{Diagrams contributing to the nucleon three-point
    correlation function, connected (left) and disconnected (right).}
  \label{fig:diagram}
  \vspace{-3ex}  
\end{figure}

The three-point correlation function has quark-connected and
quark-disconnected contributions as shown in
Fig.~\ref{fig:diagram}. Here we compute the isovector ($V$) and
isoscalar ($S$) combinations of the operator:
$j^{\mu}_{\genfrac{}{}{0pt}{}{S}{V}}(x)=\bar{u}(x)\gamma^\mu
u(x)\pm\bar{d}(x)\gamma^\mu d(x)$. Assuming SU(2) flavour symmetry the
disconnected contribution cancels for the isovector current, but does
not in the isoscalar combination. Such disconnected contributions are
exceptionally difficult to compute on the lattice due to the increased
susceptibility they exhibit to statistical fluctuations. However,
recent results using new techniques have shown that at low momentum
transfer they are small, of order of 1~\%, compared to the connected,
at pion masses of around $300$ to $400$~MeV~\cite{Alexandrou:2013wca,
  Meinel:2014}. From the isoscalar and isovector combinations we can
obtain the proton and neutron form factors through the linear
combinations: $F^p-F^n=F^u-F^d$ and $F^p+F^n=\frac{1}{3}(F^u+F^d)$.

 \hspace{-2em}  
\begin{minipage}[t]{0.6\linewidth}
  \vspace{0pt}
  \center\includegraphics[width=0.9\linewidth]{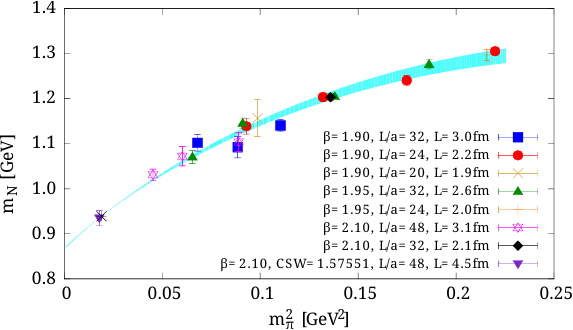}
  \captionof{figure}{The lattice spacing is set using the nucleon
    mass.}
  \label{fig:spacing}
  \vspace{1ex}  
\end{minipage}
\hfill
\begin{minipage}[t]{0.35\linewidth}
  \vspace{0pt}
  \begin{tabular}{cccc}
    \hline\hline
    \multirow{2}{*}{$\beta$} & \multirow{2}{*}{$L^3\times T$} & $a$ & $m_{\pi}$\\
                             &                                & [fm]& [GeV] \\
    \hline
    \multicolumn{4}{c}{$N_{\rm f}=2+1+1$, $c_{\rm SW}=0$}\\
    1.95 & 32$^3\times$64 & 0.082 & 0.373\\
    2.10 & 48$^3\times$96 & 0.064 & 0.210\\\hline
    \multicolumn{4}{c}{$N_{\rm f}=2$, $c_{\rm SW}=1.57551$}\\
    2.10 & 48$^3\times$96 & 0.094 & 0.130 \\
    \hline\hline
  \end{tabular}
  \captionof{table}{The twisted mass fermion ensembles used in this
    work.}
  \label{table:configs}
\end{minipage}

In this work we present three ensembles of TMF configurations; two
$N_{\rm f}=2+1+1$ ensembles with $m_\pi= 373$~MeV (referred to as
``B55'') and 210~MeV (referred to as ``D15'') and one $N_{\rm f}=2$
ensemble with $m_\pi=130$~MeV~\cite{Abdel-Rehim:2013yaa}, referred to
as the \textit{physical point} ensemble. More details are given in
Table~\ref{table:configs}. The lattice spacings are determined by
fitting the nucleon masses of 18 TMF ensembles using the leading order
heavy baryon chiral perturbation theory expansion: $m_N = m_N^0 - 4
c_1 m_\pi^2 - \frac{3 g_A^2}{16\pi f_\pi^2} m_\pi^3$, as shown in
Fig.~\ref{fig:spacing}. We require the curve to reproduce the
physical nucleon mass and allow the spacings to vary as fit
parameters. This yields the spacings listed in
Table~\ref{table:configs} for the ensembles used here.

\begin{figure}[h]
  \includegraphics[width=0.5\linewidth]{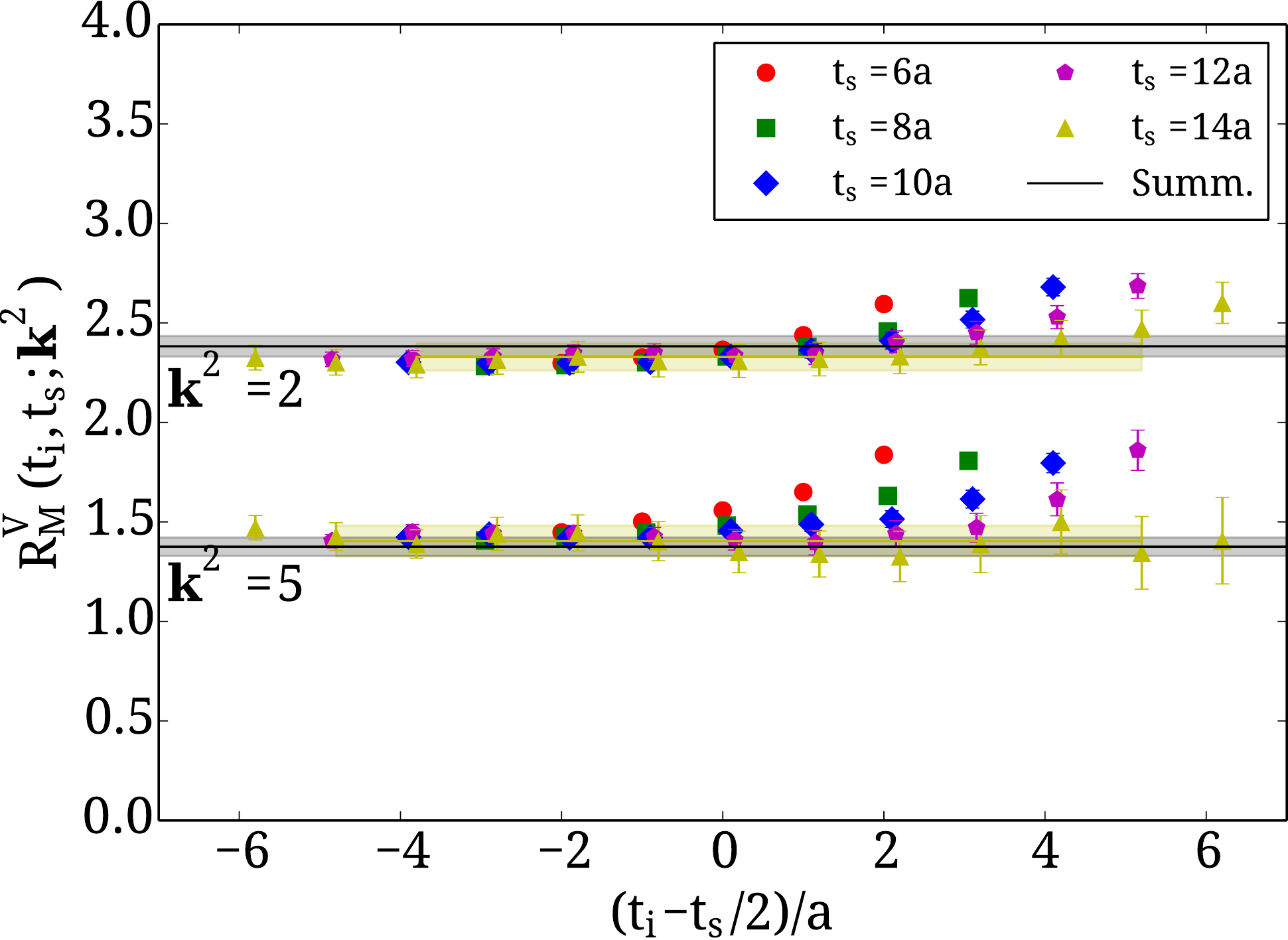}
  \includegraphics[width=0.5\linewidth]{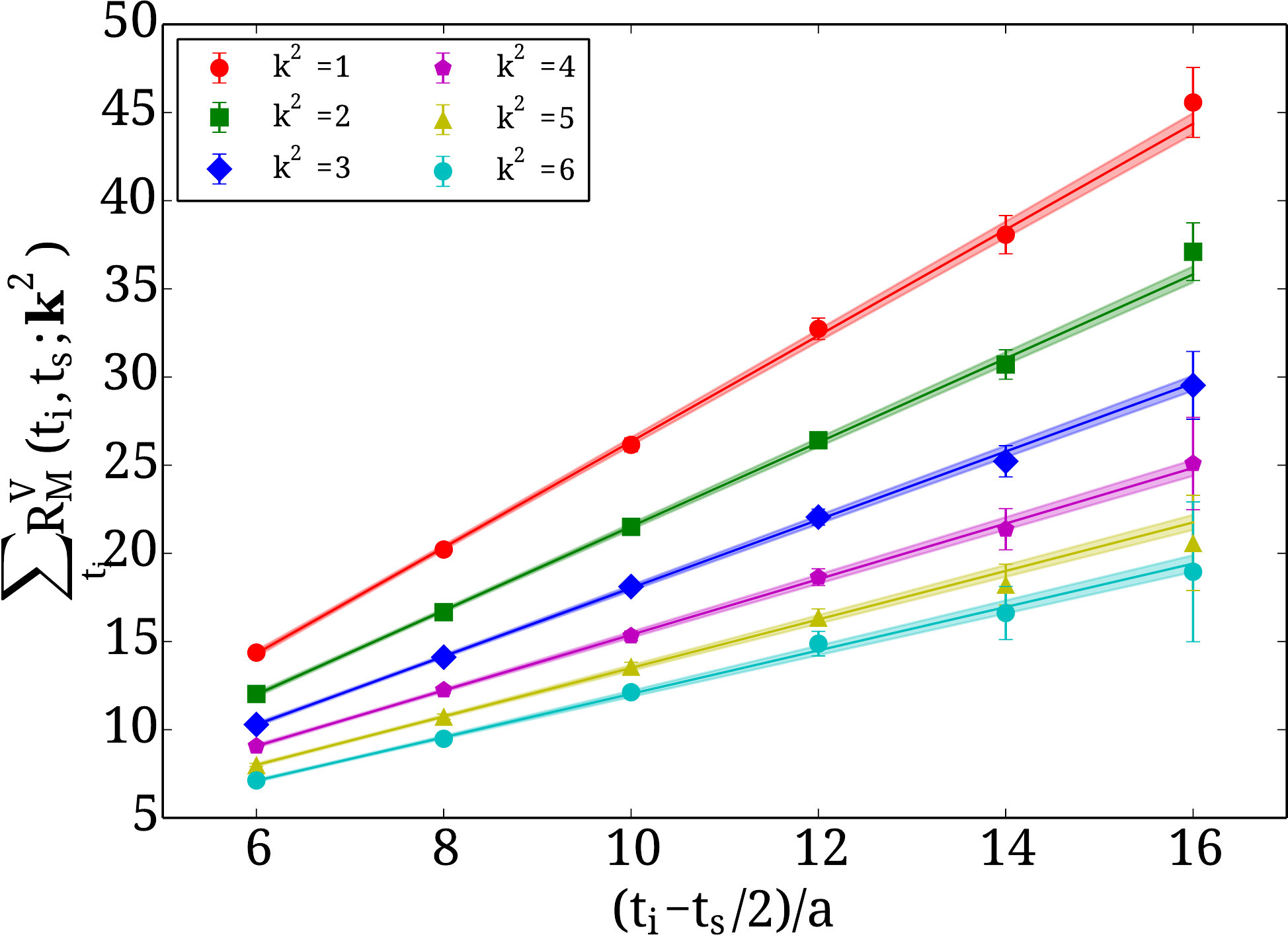}
  \caption{Extraction of the required matrix element is done via
    fitting to the plateau method (left) or the summation method
    (right). In the left panel we compare the result of the summation
    method (black line) with that of the plateau (yellow line) for the
    B55 ensemble.}
  \label{fig:ffextract}
  \vspace{-2ex}
\end{figure}

For extracting the matrix element from the lattice data, we form an
appropriate ratio of three-point to two-point functions, which cancels
unknown overlaps and exponential energy factors such that in the limit
of large time separations: $t_i-t_0\gg$ and $t_s-t_i\gg$, the
contributions from excited nucleon states are damped out, leaving a
time-independent ratio (referred to as the \textit{plateau region}):
\begin{equation}
R^V_M(t_i,t_s;\mathbf{k}^2)\xrightarrow[t_i-t_0\gg]{t_s-t_i\gg}
G^V_M(\mathbf{k}^2)[1 + O(e^{-\Delta M (t_s-t_i)},e^{-\Delta E(\mathbf{k}) (t_i-t_0})]
\end{equation}
as illustrated for the isovector magnetic form factor in
Fig.~\ref{fig:ffextract}. $\Delta M$ ($\Delta E$) is the mass (energy)
gap of the nucleon ground to nucleon first exited state. With fixed
sink time-slice, we vary the insertion time-slice to identify and fit
to the plateau region. This will be referred to as the \textit{plateau
  method}. To ensure ground state dominance, we invert for multiple
sink-source separations $t_s-t_0$, with the incremental EigCG
algorithm as an efficient multiple right-hand-side
solver~\cite{Abdel-Rehim:2013wlz}. With the availability of multiple
sink-source separations, we can also sum over the insertion time-slice
$t_i$ to obtained the summed ratio as a function of $t_s$:
\begin{equation}
\sum_{t_i}R^V_M(t_i,t_s;\mathbf{k}^2)\xrightarrow{t_s\gg}
C+G^V_M(\mathbf{k}^2)t_s[1 + O(e^{-\Delta M (t_s-t_0)},e^{-\Delta E(\mathbf{k}) (t_s-t_0)})].
\end{equation}
This allows an alternative extraction of the form factor via a two
parameter linear fit, the so-called \textit{summation
  method}. Compared to the plateau method, the advantage is that
excited states decay with a larger exponential suppression factor
determined by $t_s-t_0$ rather than $t_s-t_i$ or $t_i-t_0$, while the
disadvantage is that a two parameter fit is required. In
Fig.~\ref{fig:ffextract} we show an example of fits to the summed
ratio for momenta up to $\mathbf{k}^2=6$.

\section{Results}
\label{sec:results}

\begin{figure}[h]
  \includegraphics[width=0.5\linewidth]{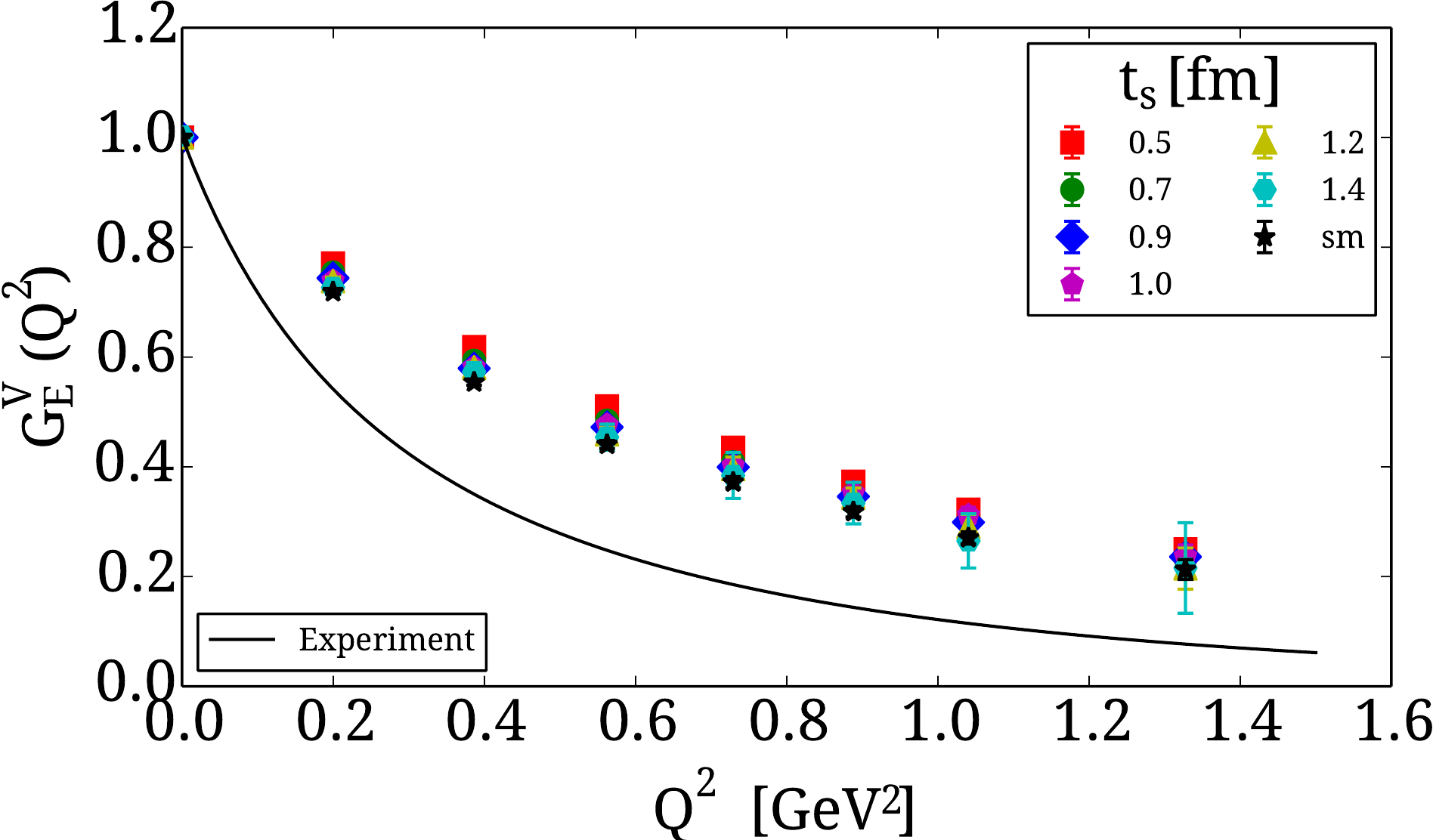}
  \includegraphics[width=0.5\linewidth]{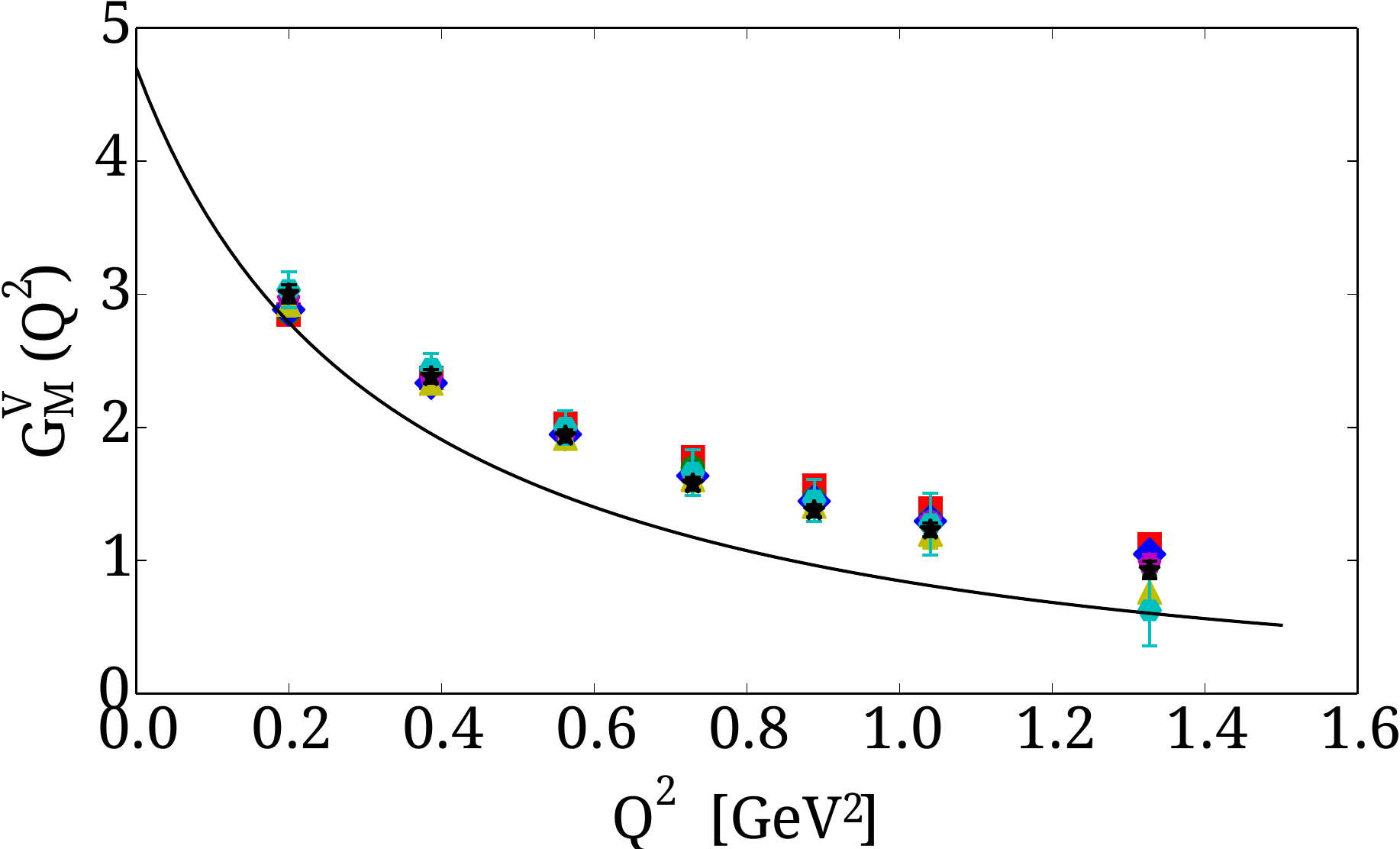}
  \caption{The electric (left) and magnetic (right) nucleon form
    factors versus the momentum transfer squared, for B55 for various
    values of $t_s$ (coloured symbols) and for the summation method
    (sm, black asterisk). The solid line is J. Kelly's
    parameterisation to the experimental data.}
  \label{fig:gegm_b55}
  \vspace{-1ex}  
\end{figure}

In Fig.~\ref{fig:gegm_b55} we show results for the B55 ensemble for
which we have carried out inversions at seven sink-source separations
for 1,200 configurations. We show the electric and magnetic isovector
Sachs form factors obtained via the plateau method and compare them to
the summation method to identify excited state effects. As the
separation increases from 0.5 to 1.4~fm the form factors become
steeper, yielding larger electric and magnetic radii. However these
results are still in tension with experiment, shown by the curve which
represents J. Kelly's parameterisation of the experimental results.

The radii are determined from the derivative of the form factors at
$Q^2=0$. Fitting the form factors to a dipole form $F_i(Q^2)=F_i(0) /
[1+Q^2/M^2_i]^2$ the radii are determined from the dipole mass via
$\langle r_i^2\rangle = 12/M_i^2$ with $i=1,\,2$, and similarly for
the Sachs form factors to obtain the electric and magnetic radii. For
the Dirac and electric form factors $F_1(0)=G_E(0)=1$ since we are
using the conserved current and thus only the mass is a fit
parameter. $F_2(0)$ is fitted for and yields the anomalous magnetic
moment of the nucleon.

\begin{figure}[h]
  \includegraphics[width=0.330\linewidth]{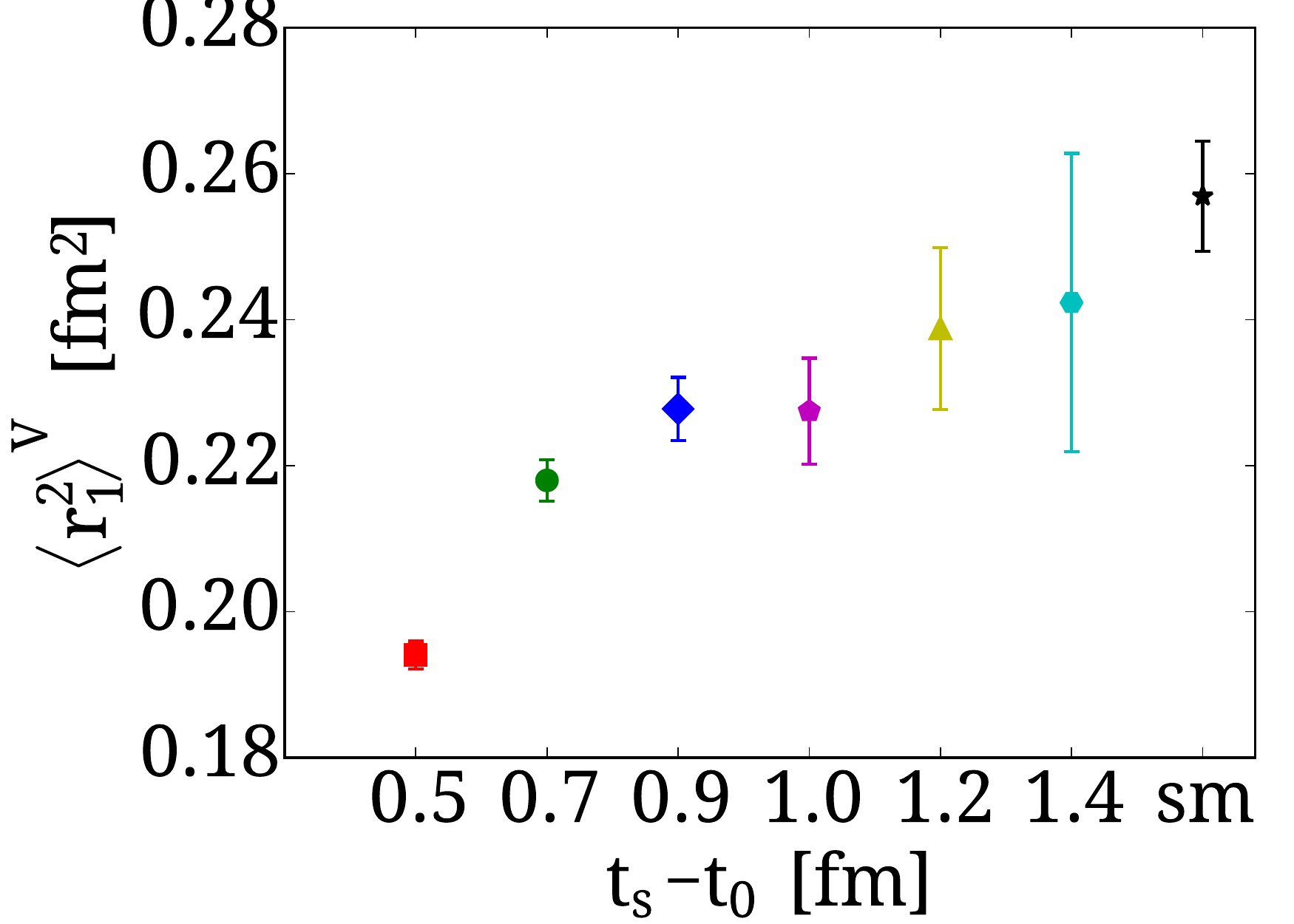}
  \includegraphics[width=0.330\linewidth]{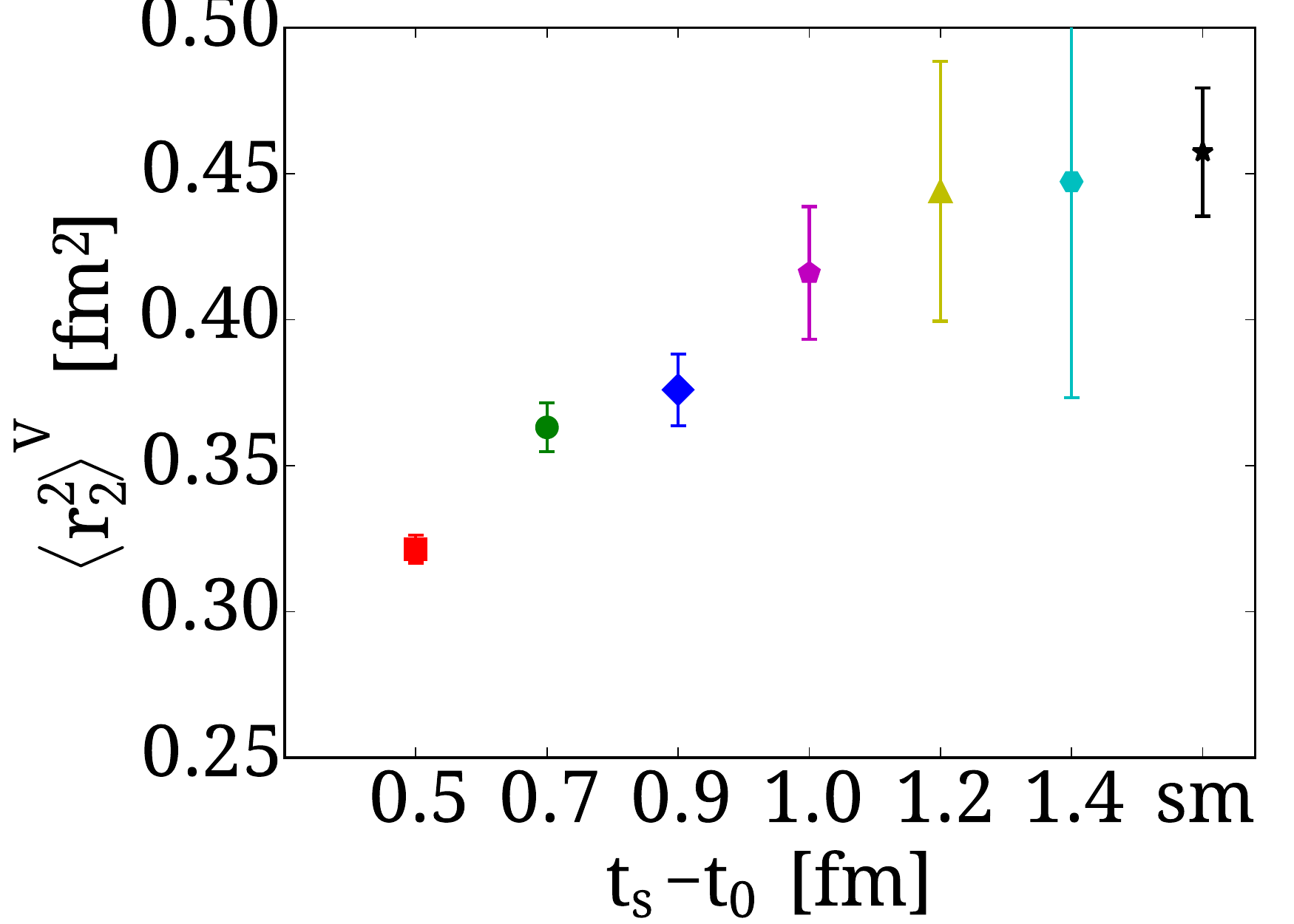}
  \includegraphics[width=0.330\linewidth]{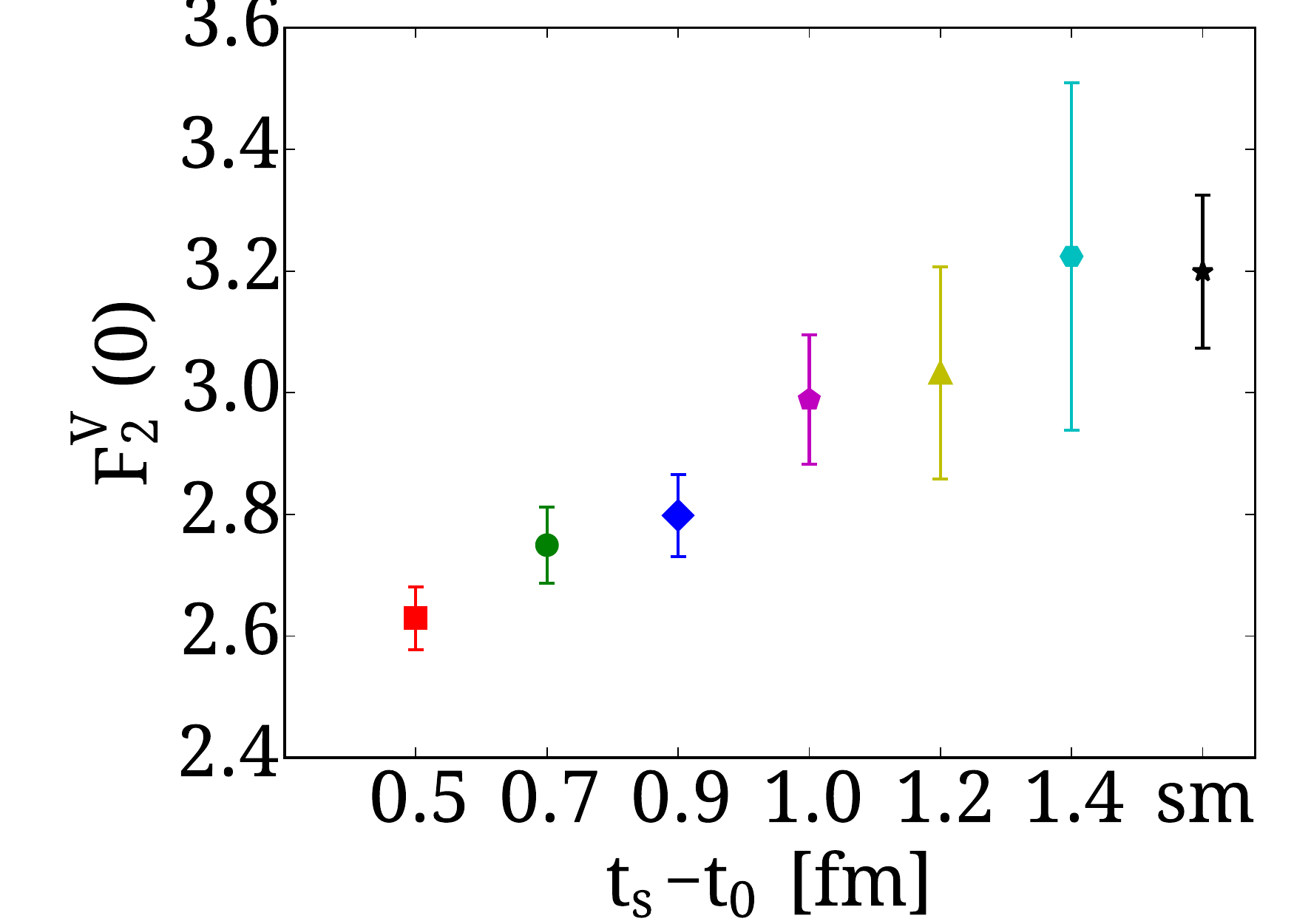}
  \caption{The Dirac and Pauli isovector radii (left and centre), and
    the isovector magnetic moment (right) for various $t_s$ and for
    the summation method. The notation is the same as in
    Fig.~\protect\ref{fig:gegm_b55}.}
  \label{fig:radii_b55}
\end{figure}

In Fig.~\ref{fig:radii_b55}, we show the isovector Dirac and Pauli
radii and the isovector anomalous magnetic moment for the B55
ensemble. Results extracted via fits to the dipole form as a function
of the sink-source separations, as well as the value extracted from
the summation method. These results corroborate the conclusion that as
$t_s$ increases the radii become larger. We also observe that
contamination from excited states are suppressed after a sink-source
separation of around 1.2-1.3~fm, indicated by the agreement between
plateau and summation methods beyond this distance.

\begin{figure}[h]
    \vspace{0pt}
    \center
    \hspace{-0.85em}
    \includegraphics[width=0.345\linewidth]{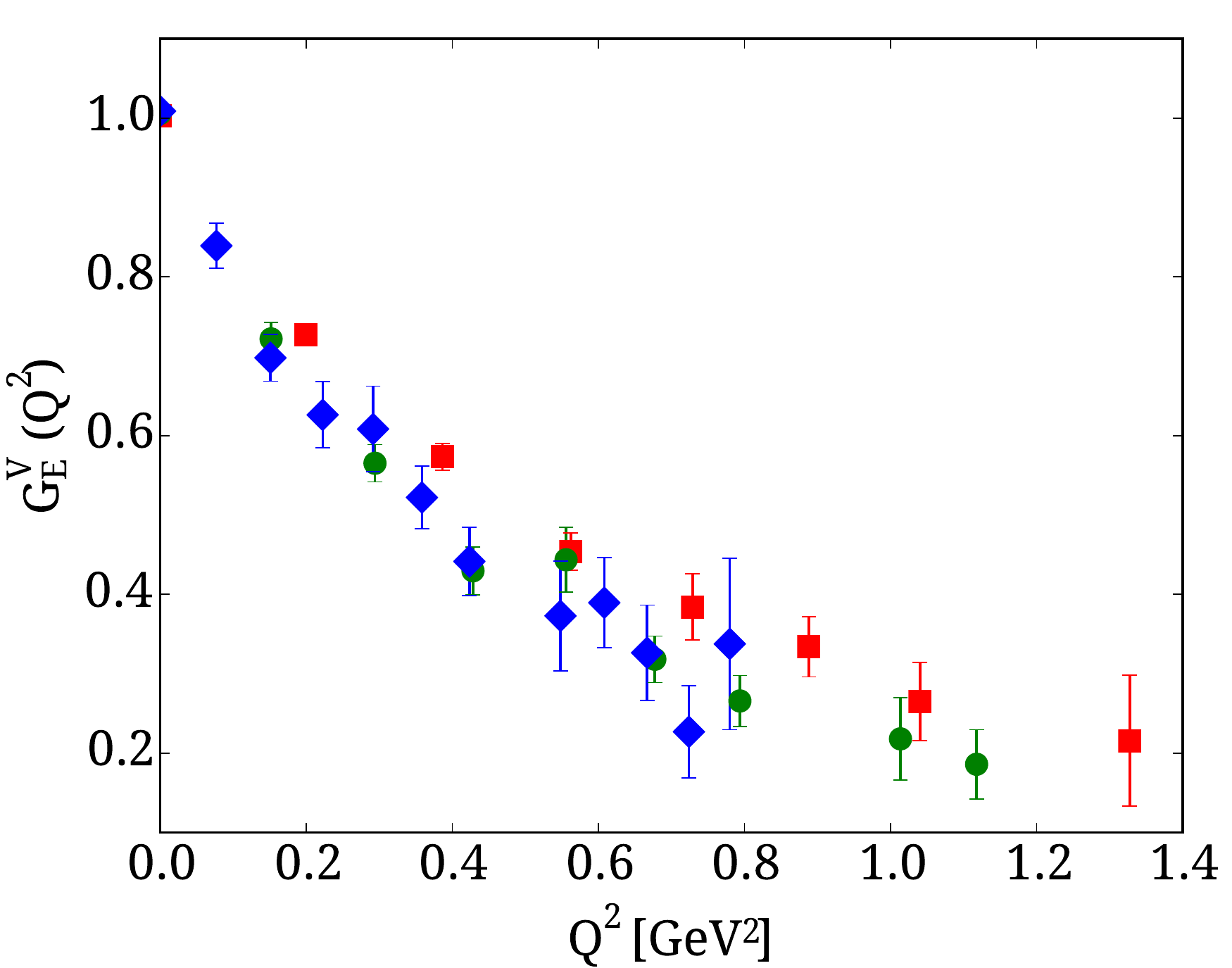}
    \hspace{-0.85em}
    \includegraphics[width=0.345\linewidth]{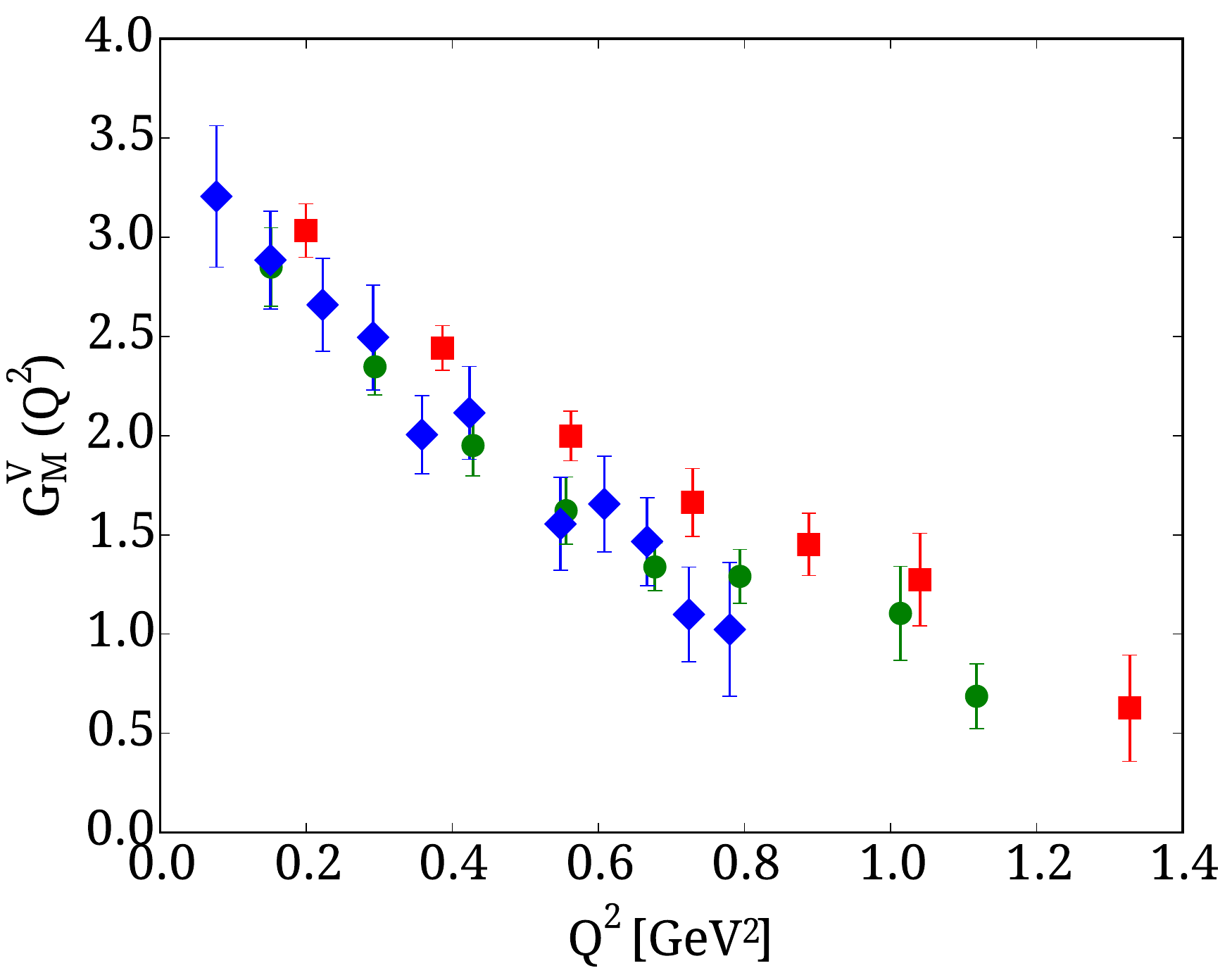}
    \hspace{-0.85em}
    \includegraphics[width=0.345\linewidth]{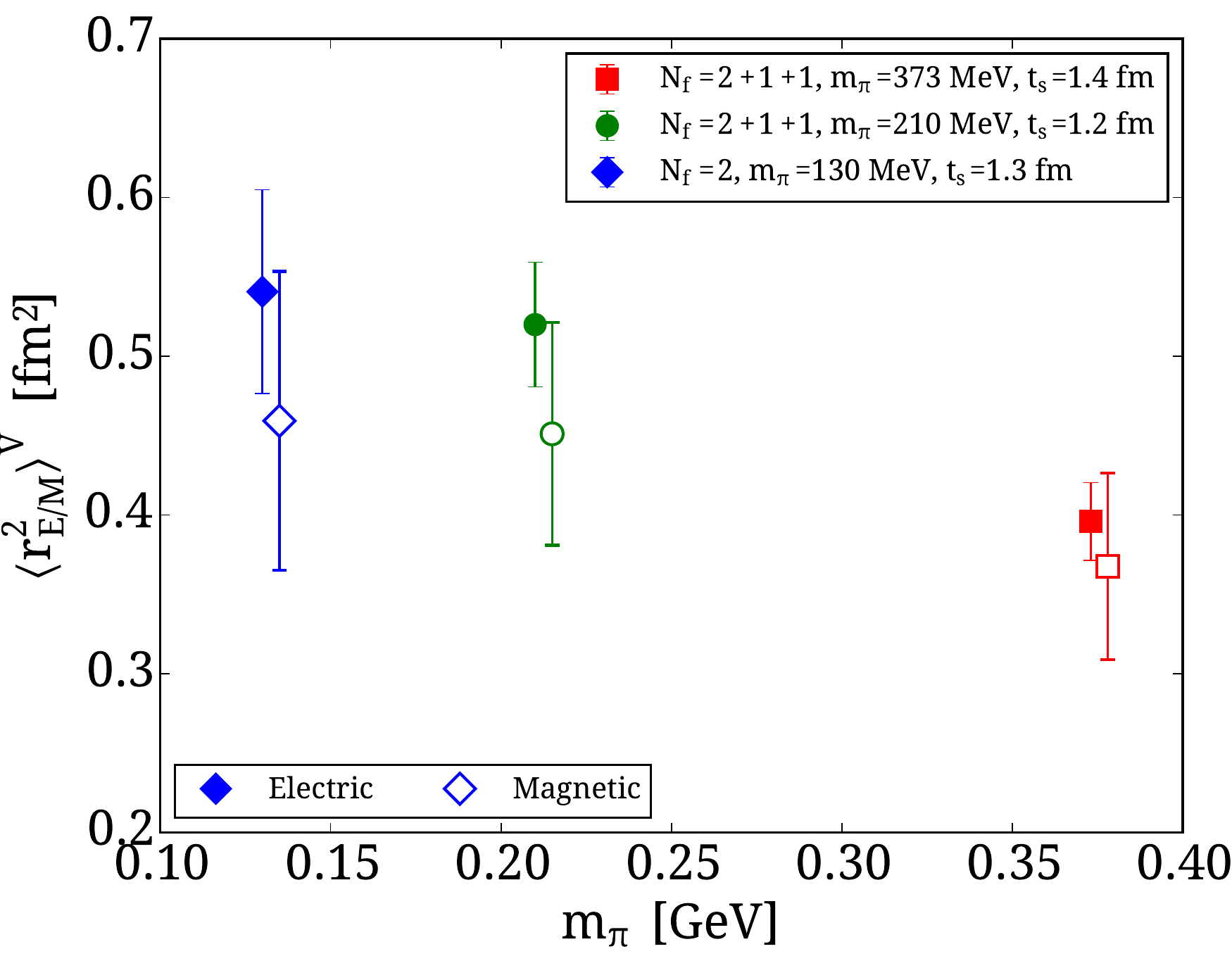}    
    \caption{The isovector electric (upper left) and magnetic (lower
      left) form factors for the three TMF ensembles of this. The
      associated radii are shown in the right panel as a function of
      the pion mass.}\label{fig:tmall}
\end{figure}

Our results using all three ensembles, including the physical point
ensemble, are shown in Fig.~\ref{fig:tmall}, for the isovector
electric and magnetic Sachs form factors. A single sink-source
separation of 1.2~fm is available for the $m_\pi$=210~MeV ensemble,
while for the physical point we have sink-source separations at 0.94,
1.13 and 1.32~fm, the largest of which is shown in
Fig.~\ref{fig:tmall}, for around 1,400 configurations. We observe a
steeper form factor as the pion mass is reduced, although more
statistics are required at the physical point for a definite
conclusion. In the same figure the electric and magnetic radii are
also shown as a function of the pion mass, confirming the tendency
towards larger values as the pion mass is decreased.

\begin{figure}[h]
  \hspace{-0.85em}
  \includegraphics[width=0.345\linewidth]{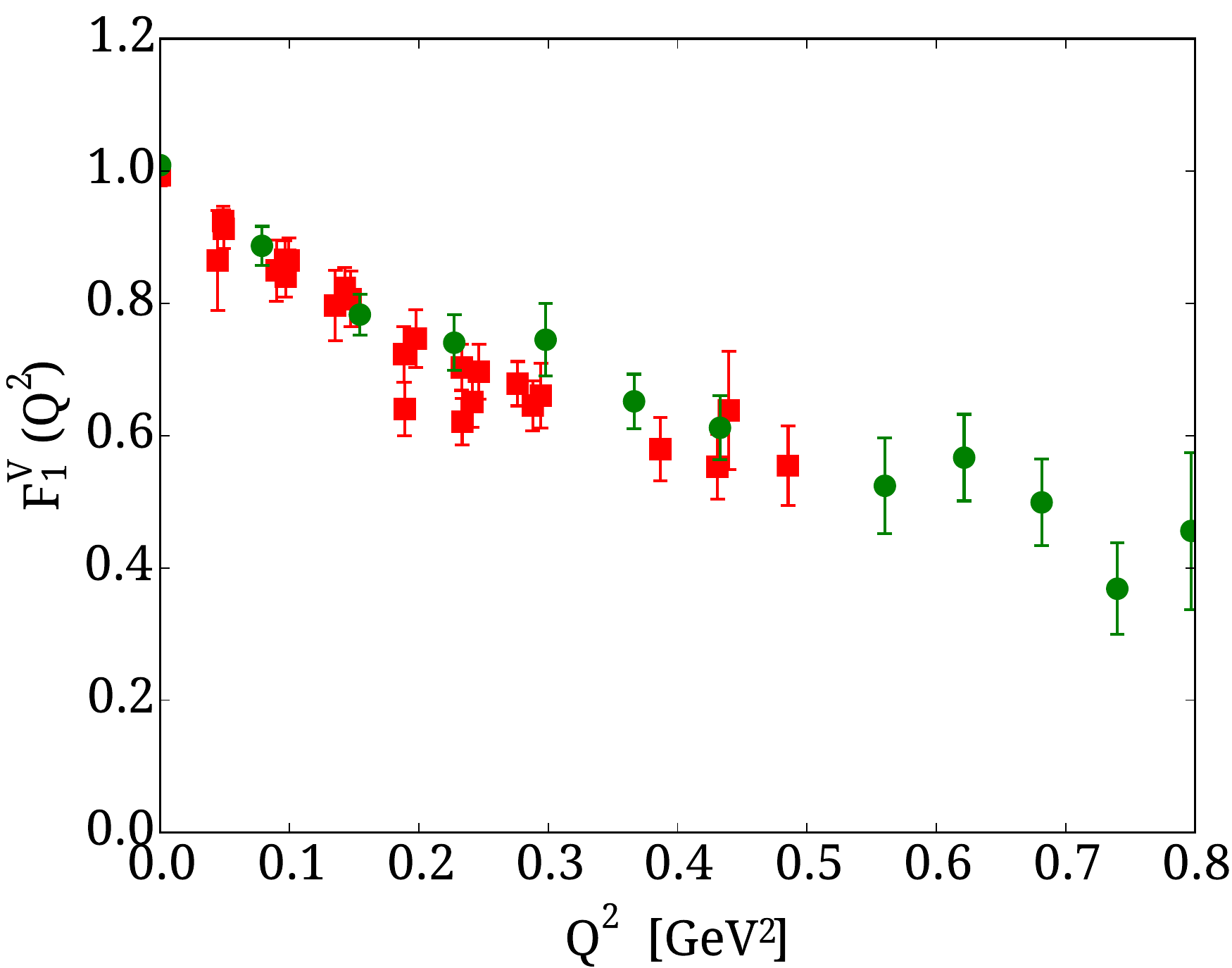}
  \hspace{-0.85em}
  \includegraphics[width=0.345\linewidth]{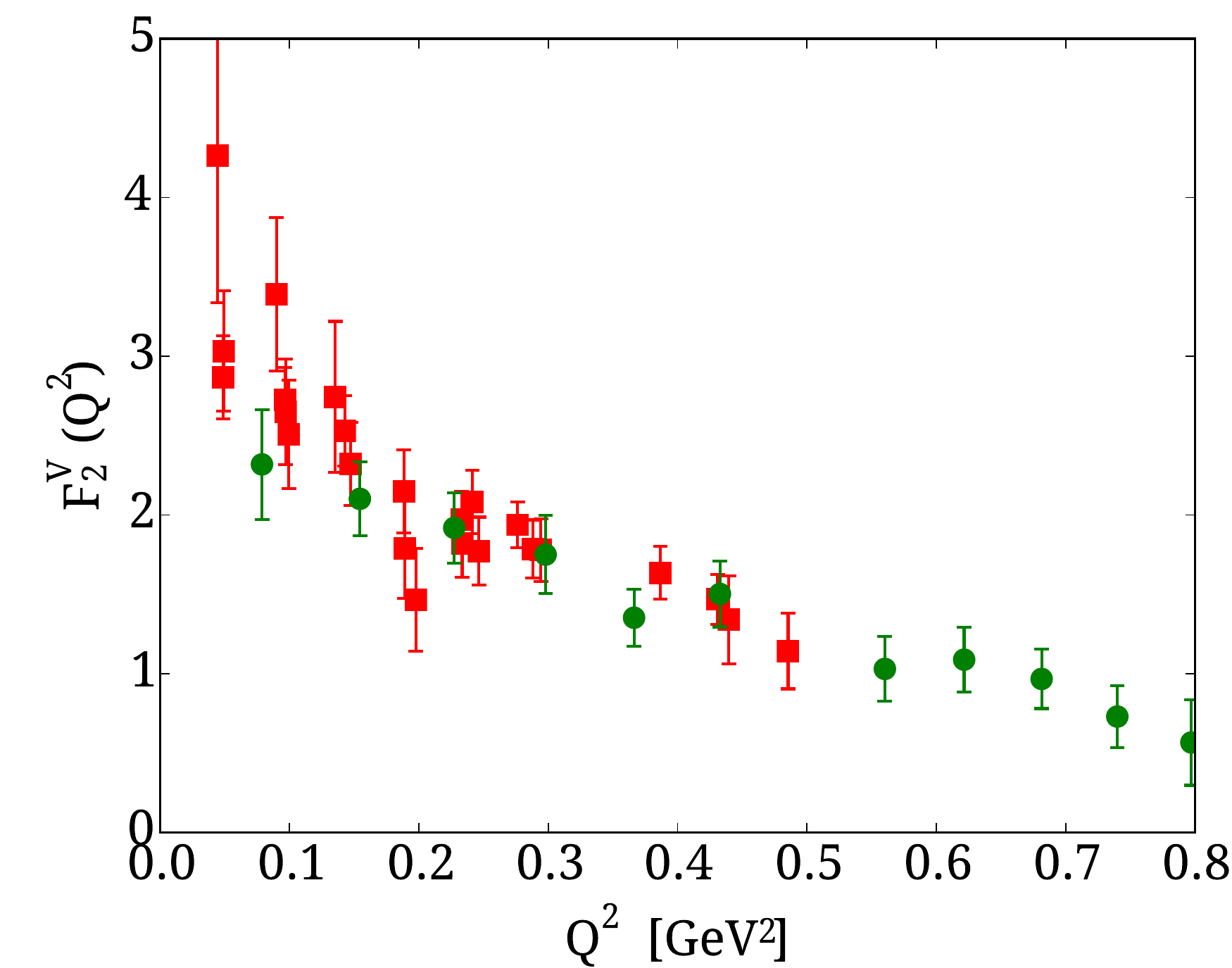}
  \hspace{-0.85em}
  \includegraphics[width=0.345\linewidth]{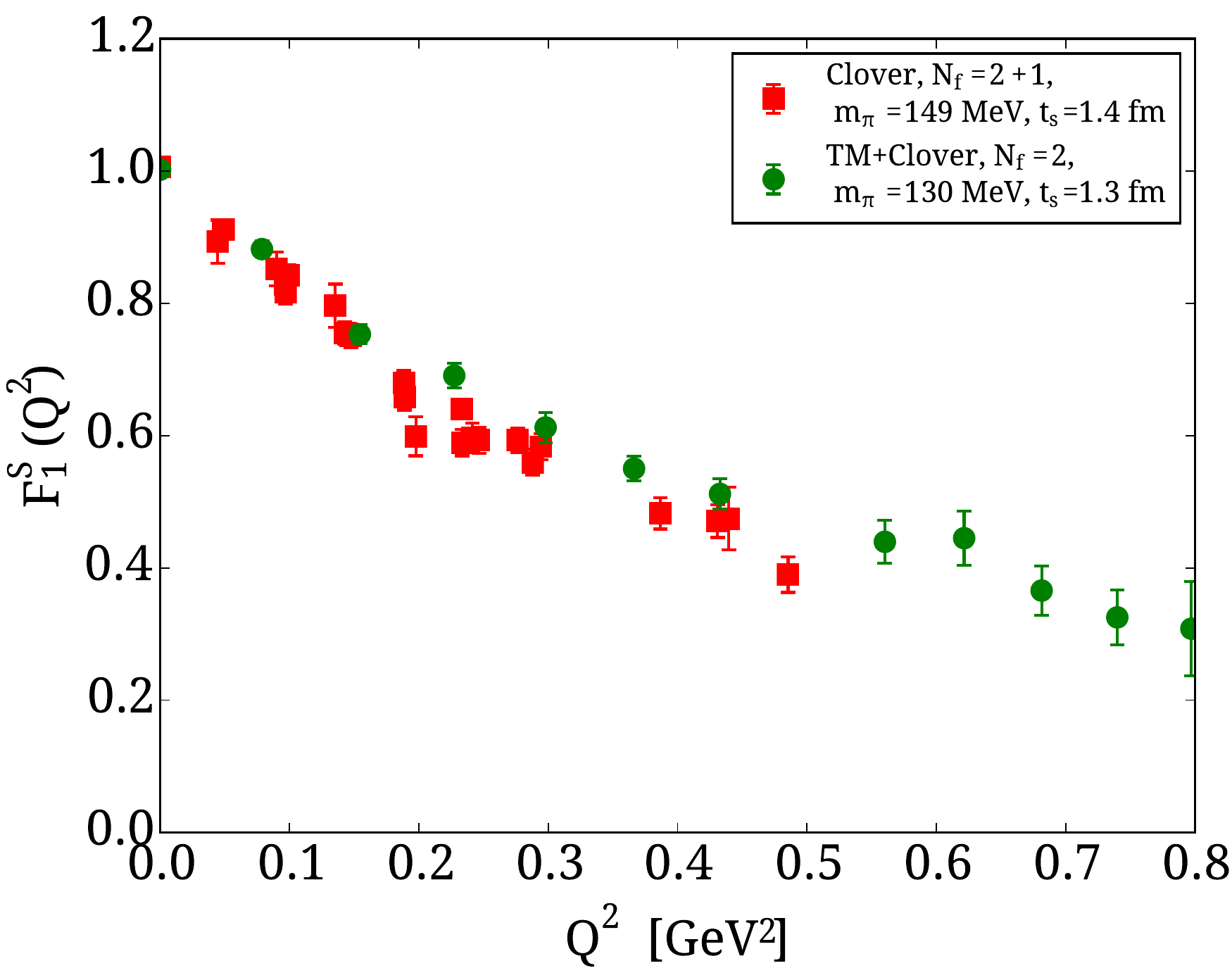}
  \caption{The isovector Dirac (left) and Pauli (centre), and the
    isoscalar Dirac (right) form factors of this work, compared to
    results from the LHPC at
    $m_\pi$=149~MeV~\protect\cite{Green:2014xba}.}\label{fig:lhpc}
\end{figure}

We compare our results at the physical point with those of a recent
calculation using clover fermions at similar pion mass in
Fig.~\ref{fig:lhpc}~\cite{Green:2014xba}. We see a consistency in the
extracted form factors between the two formulations. The isoscalar
Pauli form factor $F^S_2$, not shown in Fig.~\ref{fig:lhpc}, is found
consistent with zero in both cases.

\begin{figure}[h]
  \includegraphics[width=0.5\linewidth]{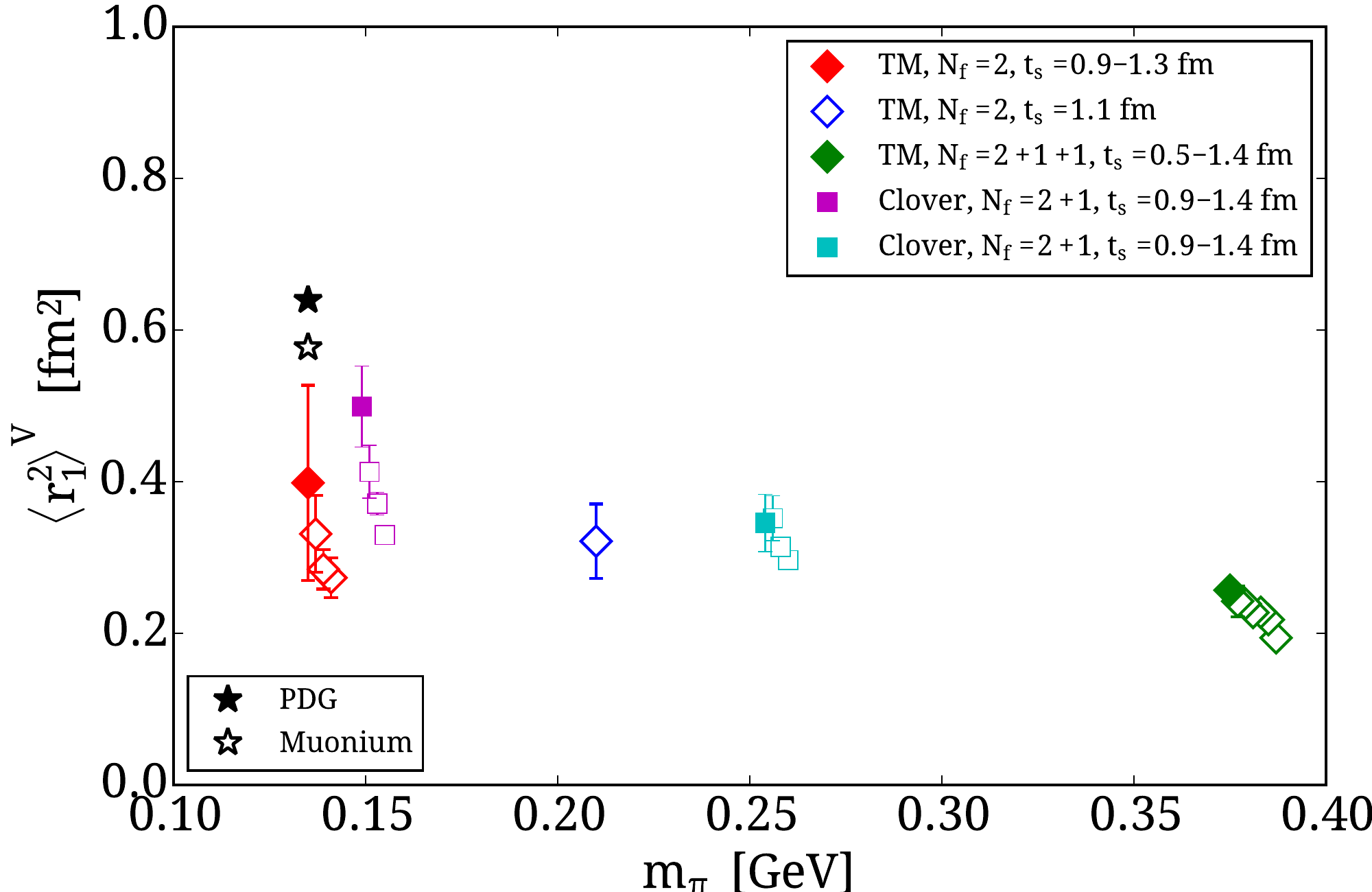}
  \includegraphics[width=0.5\linewidth]{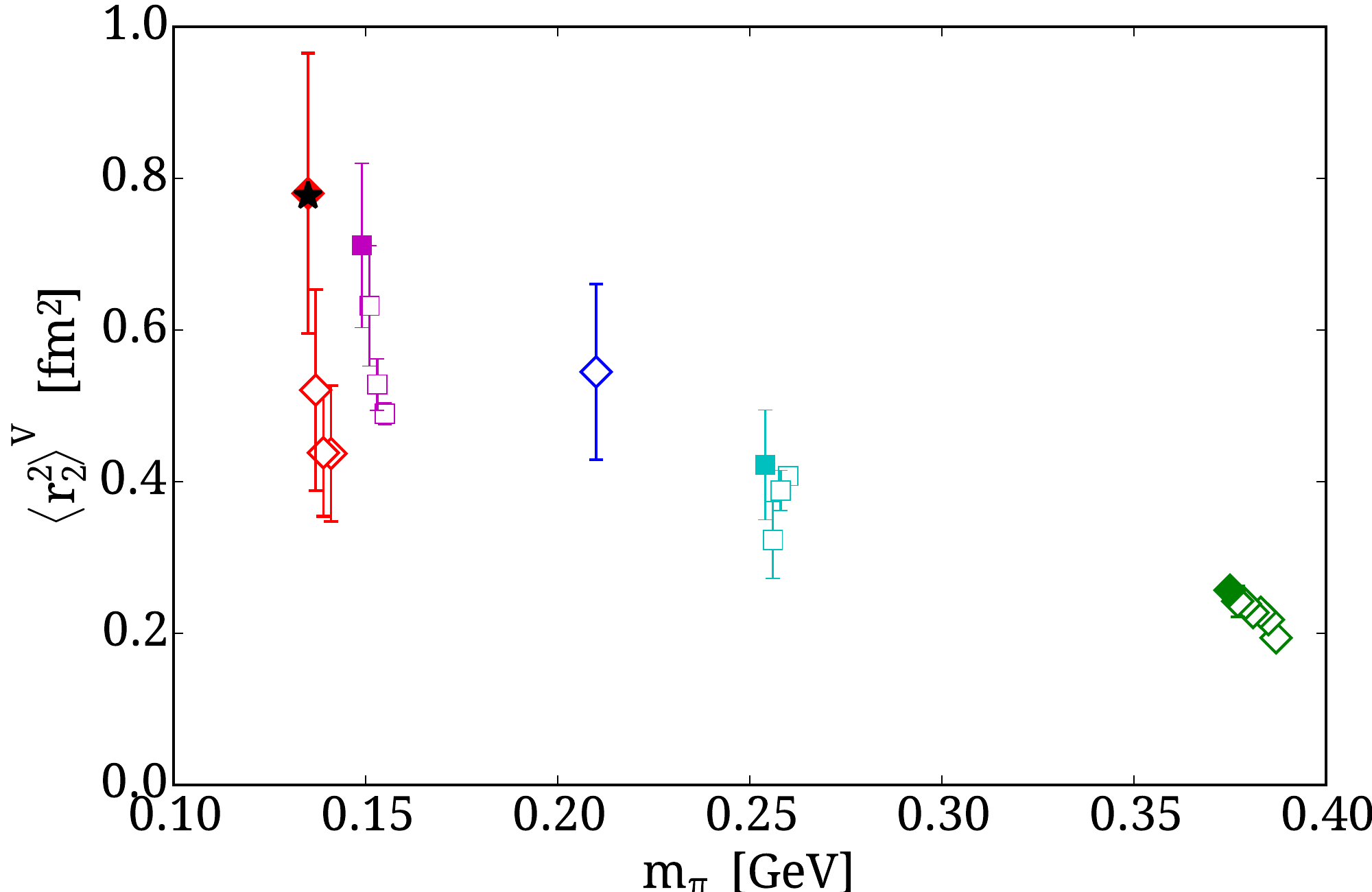}
  \caption{The isovector Dirac (left) and Pauli (right) radii as a
    function of the pion mass. The diamonds are from this work while
    the squares are from the LHPC. Filled symbols are results using
    the summation method, while open symbols are from the plateau
    method.}
  \label{fig:radii_lhpc}
\end{figure}

In Fig.~\ref{fig:radii_lhpc} we show the isovector Dirac and Pauli
radii as a function of the pion mass, for the TMF ensembles analysed
in this work and the clover ensembles analysed in
Ref.~\cite{Green:2014xba}. We also show the PDG values of these
quantities and, for the case of $\langle r_1^2\rangle^V$, we include
the recent result obtained from muonium Lamb shift measurements. As
the pion mass decreases the effect of the excited states seem more
pronounced indicated by the spread of the open symbols. However, the
errors also become larger and a study with increased accuracy is
required. Both clover and TMF results are in agreement, and seem to be
converging towards the experimental values, although errors are large
and need to be drastically reduced to make contact with experiment.

\section{Conclusions}
\label{sec:conclusions}
We have presented preliminary results for the electromagnetic form
factors of the nucleon including a lattice QCD ensemble with a pion
mass set to its physical value. A lattice at a heavier pion mass of
373~MeV was used for a thorough study of excited state effects,
indicating that a three-point function sink-source separation of
$\sim$1.3~fm is sufficient for ensuring ground state dominance. Our
results at the physical point are in agreement with recent results
using clover fermions at a similar pion mass. Within large statistical
errors, our results are either in agreement or tend towards the
experimental values as the pion mass is reduced. In order to make
contact with experiment, a 2\% error in the radii is required, which
amounts to more than a 10-fold increase in statistics. Deflation
algorithms in combination with
all-mode-averaging~\cite{Shintani:2014vja} are currently under
investigation to efficiently achieve this.

\vspace{1ex}
\noindent\textbf{Acknowledgements:} M.C., K.H. and K.J. acknowledge
support by the Cyprus RPF under contracts
TECHNOLOGY/$\Theta$E$\Pi$I$\Sigma$/0311(BE)/16,
T$\Pi$E/$\Pi$$\Lambda$HPO/0311(BIE)/09 and
$\Pi$PO$\Sigma$E$\Lambda$KY$\Sigma$H/ EM$\Pi$EIPO$\Sigma$/0311/16
respectively. This project used computer time granted by NIC on
JUQUEEN (project hch02) and JUROPA (project ecy00) at the JSC as well
as by the Cyprus Institute on (project lspro113), under the Cy-Tera
project (NEA Y$\Pi$O$\Delta$OMH/$\Sigma$TPATH/0308/31).
\vspace{-2ex}
\setlength\bibitemsep{-0.2\itemsep}
\printbibliography
\end{document}